\documentclass[notitlepage,12pt]{article}

\usepackage{graphicx} 
\usepackage{amssymb}
\usepackage{color}

\usepackage{fullpage}

\def\thetacirc{\stackrel{\,\,\circ}{\theta}}
\def\thetabullet{\stackrel{\,\,\bullet}{\theta}}
\def\xcirc{\stackrel{\,\,\circ}{x}}
\def\xbullet{\stackrel{\,\,\bullet}{x}}

\def\rcirc{\stackrel{\,\,\circ}{r}}
\def\rbullet{\stackrel{\,\,\bullet}{r}}
\def\ucirc{\stackrel{\,\,\circ}{u}}
\def\ubullet{\stackrel{\,\,\bullet}{u}}
\def\psic{\stackrel{\,\,\circ}{\psi}\!}
\def\psib{\stackrel{\,\,\bullet}{\psi}\!}
\def\Deltac{\stackrel{\,\,\circ}{\Delta}\!}
\def\Deltab{\stackrel{\,\,\bullet}{\Delta}\!}
\def\Hcirc{{\cal H}^{(\circ)}}
\def\Hbullet{{\cal H}^{(\bullet)}}
\def\Acirc{{\hat A}^{(\circ)}}
\def\Abullet{{\hat A}^{(\bullet)}}

\def\graphwidth{5cm}
\def\graphsep{0.5cm}

\title{Coherent States for infinite homogeneous waveguide arrays: Cauchy coherent states for $E(2)$}

\begin{document}

\author{Julio Guerrero  \thanks{ Department of Mathematics, University of Ja\'en, Campus Las Lagunillas s/n, 23071 Ja\'en, Spain \texttt{jguerrer@ujaen.es: corresponding author} } \thanks{Institute Carlos I of Theoretical and Computational Physics (iC1), University of  Granada,
  Fuentenueva s/n, 18071 Granada, Spain. } \and Francisco F. L\'opez-Ruiz \thanks{Department of Applied Physics, University of C\'adiz, Campus de Puerto Real, E-11510 Puerto Real, Cádiz, Spain. 
 \texttt{paco.lopezruiz@uca.es} }}



\date{\today}

\begin{abstract}
 Perelomov coherent states for equally spaced, infinite homogeneous waveguide arrays with Euclidean E(2) symmetry are defined, and a new resolution of the identity is obtained. The key point to construct this novel resolution of the identity is the fact that coherent states satisfy the Helmholtz equation (in coherent states labels), and thus every coherent state belongs to a one-parameter family uniquely determined by the Cauchy initial data of the coherent state in a one-dimensional Cauchy set. For this reason we call \textit{Cauchy coherent} states to these initial data. The novel, non-local resolution of the identity in terms of Cauchy coherent states is provided using frame theory. It is also shown that Perelomov coherent states for the Eucliean E(2) group have a simple and natural physical realization in these waveguide arrays.
\end{abstract}



\maketitle

\section{Introduction}

Waveguides (see, for instance, \cite{Jones65}) are optical devices made  of optical fibres, i.e. an infinite cylinder of dielectric material (the core, with a high index of refraction) surrounded by another material (the cladding, with a lower index of refraction). They are built up in such a way that they guide light in the core (by total reflection on the cladding) in the longitudinal direction, whereas the transverse  size of the core is of the order of magnitude of the wavelength of the light. An array of waveguides is a set of waveguides in such a way that light couples between nearest waveguides by evanescent fields \cite{Jackson,Jones65}. We shall restrict ourselves to the case of 1D parallel waveguides containing a single guided mode, which are easier to describe.

The simplest parallel waveguide array  is that with equally spaced waveguides and of homogeneous properties in the longitudinal direction. And the simplest among them is the case of an infinite array, since it possesses both (continuous) translational
symmetry  in longitudinal direction and (discrete) translational symmetry in the transverse direction. We shall see later that these 1D infinite arrays possess the symmetry of the Euclidean E(2) group. The cases of semi-infinite and finite waveguide arrays are considered in \cite{Guerrero21b}.

We shall show that the differential equations describing the amplitude of the electric field in each waveguide for the propagation along the longitudinal direction can be expressed in terms of the generators of the Euclidean E(2) group, known as (true)  phase operators \cite{Louisell63,Newton80}, which are naturally unitary and act on an extended Fock space (obtained by adding to the ordinary Fock number states  the negative number states). Note that the same group E(2) is the symmetry of the quantum mechanics of a particle on the circle  \cite{Ohnuki93}. We shall see that in fact both system are intimately related.

Since the Euclidean groups do not possess  square integrable representations \cite{DeBievre89,Isham91},  coherent states for the group $E(2)$ (and the higher dimensional extensions $E(n)$) of the Perelomov type \cite{Perelomov86}  do not admit a resolution of  the identity. Diverse techniques have been devised to circumvent this problem, like in  \cite{DeBievre89}, where the labels of the coherent states are restricted to the cylinder and further admissibility conditions are required on the fiducial or vacuum state (see also \cite{Torresani95}
for applications on signal processing on the circle). See \cite{Leon07,Fresneda18} for a recent account of these states. In \cite{Isham91},  a reducible representation of E(2) is used summing up all irreducible representations with the  radius of the circle belonging to an interval (considering in fact a label space defined by an annuls times a line). 
There are other approaches for the definition of coherent states on the circle, which are not related to the E(2) group, like \cite{DeBievre93a} or  \cite{Gonzalez98}. See the review \cite{Kastrup06}  for details on these and
other families of coherent states on the circle.

In this paper we shall address the issue of constructing a new  resolution of the identity for coherent states of the E(2) group (in the flavor of functions representing configurations of light amplitudes  on an infinite waveguide array) modifying the usual integration on the phase space (the coadjoint orbit of the group) with an invariant measure under E(2) by a double integral with a convolution kernel on a non-homogeneous subspace, the one used to define a Cauchy initial value problem for the Helmholtz equation \cite{Wolf81}, giving rise to the notion of Cauchy coherent states. The Helmholtz equation appears here as the eigenvalue equation of the Casimir of E(2) and has nothing to do with the fact that we are describing an optical system, being rather a consequence of the symmetry of the system, appearing also in other contexts like the momentum representation of a particle on the circle (see \cite{SphereMomentum20} for the case of
the sphere $S^3$, which can easily generalized to any sphere $S^n$).
This construction can also be performed in the usual representation of E(2) in terms of functions on the circle, and can be easily generalized to higher dimensional cases (see  \cite{CSE2} for details).

The content of the paper is as follows. In Section \ref{HWGA} we review the study of an infinite set of equally spaced homogeneous waveguide arrays, obtaining the differential equations describing the amplitudes of light along the waveguides and computing the propagator. In Section \ref{E(2)-CS} we recall that the symmetry of this system is the Euclidean group and we construct coherent states of the Perelomov type. In Section  \ref{ResolutionIdentity} a resolution of the identity for these coherent states is built using the fact that coherent states satisfy the Helmholtz equation. In Sec. \ref{PhysicalInterpretation} a physical realization of E(2) coherent states and of Cauchy Coherent states is provided in tilted waveguide arrays.
Finally, we present some conclusions  of the present work in Section \ref{conclusions}.

\section{Equally spaced, infinite homogeneous waveguide arrays}
\label{HWGA}

Due to its versatility, waveguide arrays have become one of the best devices to simulate both classical and quantum phenomena \cite{Mar-Sarao08,PerezLeija10,LeonMontiel11a,PerezLeija12,RodriguezLara14}, and in this case we shall use them to realize coherent states of the Euclidean E(2) group. We shall focus in  the case of an infinite number of equally spaced and homogeneous parallel waveguide arrays.

Denote by $A_n(z)$ the electric modal field at position $z$ in the $n$-th waveguide of an inifite waveguide array, and suppose that both the propagation constant and the coupling constant between adjacent waveguides are the same for the array (homogeneity). Then, the equations governing the propagation of  $A_n(z)$ are given by (see \cite{Jones65}):
\begin{equation}
 i\frac{d A_n}{d z} = A_{n+1}+A_{n-1}\,,\qquad n\in\mathbb{Z}\,,
 \label{CoupledEqinf}
\end{equation}
where we have performed a suitable transformation to remove the propagation constant and 
  $z$ is  measured in units of the inverse of the coupling constant (see \cite{Makris06}).
  
Let us introduce Dirac notation, which is more convenient in this setting (see \cite{RodriguezLara14}). Consider an abstract Hilbert space ${\cal H}$ generated by the extended Fock basis $\bar{\cal F}=\{|n\rangle,\,n\in\mathbb{Z}\}$, where the Fock state $|n\rangle$ is given by $|n\rangle=(\ldots,0,\stackrel{(n)}{1},0,\ldots)$, representing a constant (normalized) light amplitude along  the $n$-th waveguide, with zero amplitude in all other waveguides. This way, the Hilbert space ${\cal H}$ is  isomorphic to $\ell^2(\mathbb{Z})$.

The (infinite length) vector  ${\cal A}=(\ldots,A_0,\ldots)$ can thus be written  as:
\begin{equation}
  |{\cal A}\rangle=\sum_{n\in\mathbb{Z}} A_n|n\rangle\,.
\end{equation}

The Eqs. (\ref{CoupledEqinf}) adquires the form of  a Schr\"odinger equation, where the time variable is replaced by $-z$:
\begin{equation}
 -i \frac{d\ }{d z}|{\cal A}\rangle = \hat{H}  |{\cal A}\rangle\,,
\end{equation}
and the Hamiltonian is  $\hat{H} =\hat{V}^\dagger+ \hat{V}$,
where $\hat{V}^\dagger$ and $\hat{V}$ are step up and down operators, respectively:
\begin{equation}
\hat{V}|n\rangle = |n-1\rangle\,,\qquad
 \hat{V}^\dagger|n\rangle = |n+1\rangle\,,\qquad n\in\mathbb{Z}\,. \label{VVd}
\end{equation}

Vectors in ${\cal H}$ correspond to light distributions along the whole waveguide with finite total energy at each $z$:
\begin{equation}
 ||{\cal A}(z)||^2 = |||{\cal A}(z)\rangle||^2=\sum_{n\in\mathbb{Z}} |A_n(z)|^2 < \infty\,.
\end{equation}

The Hamiltonian $\hat{H}$ models, in the context of linear coupling theory \cite{Jones65}, the evanescent coupling among adyacents waveguides.

As in the case of the Schr\"odinger equation, a propagator $\hat{U}(z)$ can be introduced, $|{\cal A}(z)\rangle = \hat U(z)|{\cal A}(0)\rangle$, which is given by:
\begin{equation}
 \hat{U}(z)=e^{i z \hat{H}}=\sum_{n,m\in \mathbb{Z}} i^{n-m} J_{n-m}(2z)|n\rangle\langle m|\,,
\end{equation}
where the commutation relations of  the operators  (\ref{VVd}),  given in Eq. (\ref{commutators}), have been used, and with $J_n$ being Bessel functions of the first kind.

Let us generalize Hamiltonian $\hat H$ and introduce a  \textit{tilted} Hamiltonian\footnote{This kind of Hamiltonian describes  waveguide arrays that are tilted \cite{PhysRevLett.85.1863,PhysRevLett.88.093901}, where the the tilting can be produced either by manufacturing tilted waveguide arrays or by tilting the wavefront of the incident light in an ordinary waveguide array. For convenience, the tilting $\varphi$ has been shifted by $\frac{\pi}{2}$, $\varphi=\theta-\frac{\pi}{2}$, so the non tilted case $\varphi=0$ is recovered for $\theta=\frac{\pi}{2}$. Note that the angle $\varphi$ is related to the transversal component $k_x$ of the wave number vector $\vec{k}$ through $\varphi=k_x d$, with $d$ the separation between the waveguides. The actual (geometrical) tilting of the waveguides is given by $\alpha=\textrm{arcsin} (\frac{\varphi}{k d})=\textrm{arcsin} (\frac{k_x}{k})$, with $k$ the norm of $\vec{k}$.}:
\begin{equation}
\hat{H}^\theta = e^{i (\theta-\frac{\pi}{2} )\hat{n}}\hat{H}e^{-i (\theta -\frac{\pi}{2})\hat{n}}=  
-i (e^{i\theta} \hat{V}^\dagger-e^{-i\theta}\hat{V})\,, \label{Htilted}
\end{equation}
with $\hat{n}|n\rangle = n|n\rangle\,,\,n\in\mathbb{Z}$ the number operator. The associated \textit{tilted} propagator is given by:
\begin{equation}
 \hat{U}^\theta(z)=e^{i z \hat{H}^\theta} = e^{z( e^{i\theta} \hat{V}^\dagger-e^{-i\theta}\hat{V}   )}=\sum_{n,m\in \mathbb{Z}} e^{i(n-m)\theta} J_{n-m}(2z)|n\rangle\langle m|\,.
\end{equation}

For $\theta=\frac{\pi}{2}$ the initial Hamiltonian $\hat H$ and propagator $\hat U$ are recovered.

It should be emphasized that the Hamiltonians $\hat{H}^\theta$ are self-adjoint and therefore the propagators $\hat{U}^\theta(z)$ are unitary. This means that the total energy $||{\cal A}(z)||^2 =\sum_{n\in\mathbb{Z}} |A_n(z)|^2$ is preserved along propagation in the $z$ direction. An important consequence of this is that the total energy holded by the whole waveguide array (for all values of z) is infinite\footnote{Experimentally we are only able to build waveguide arrays with a finite number of elements and of finite length in the $z$ direction. However, with a sufficiently large number of array elements and  long enough waveguides (but holding a finite amount of total energy) we can approximately reproduce the construction of this paper, see \cite{PhysRevLett.85.1863,PhysRevLett.88.093901} for experimental realizations.}. This will be important in the understanding of some issues in next sections.

In the next Section,  we explore E(2) Perelomov-type coherent states generated by a displacement operator that matches $\hat{U}^\theta$.
In Sec. \ref{PhysicalInterpretation} we show graphically the different behaviours of the Hamiltonians $\hat{H}^\theta$ with propagators $\hat{U}^\theta$ when the input light has a Gaussian profile with different widths.

\section{E(2) coherent states}
\label{E(2)-CS}

The step operators $\hat{V}^\dagger$ and $\hat{V}$ introduced in the previous Section verify
%
the unitarity property:
\begin{equation}
 \hat{V}\hat{V}^\dagger=\hat{V}^\dagger\hat{V}=\hat I_{\cal H} \label{unitarity}
\end{equation}

They close, jointly with the number operator $\hat{n}$, the Euclidean algebra $E(2)$, with commutation relations:
\begin{eqnarray}
  \left[\hat{n},\hat{V}\right]  &=& -\hat{V} \nonumber \\
  \left[\hat{n},\hat{V}^\dagger \right] &=& \hat{V}^\dagger \label{commutators}\\
  \left[\hat{V},\hat{V}^\dagger \right] &=& 0\nonumber  \,.
\end{eqnarray}

It turns out that they constitute
a unitary and irreducible realization of $E(2)$,  the unitarity property (\ref{unitarity})  being the eigenvalue equation  for the quadratic Casimir of the Euclidean algebra $E(2)$, $\hat{C}_2=\hat{V}\hat{V}^\dagger=\hat{V}^\dagger\hat{V}$, i.e. $\hat{C}_ 2=\hat I_{\cal H}$.

We can introduce Perelomov-type coherent states by means of the Displacement operator:
\begin{equation}
 \hat D(\alpha)=e^{\alpha \hat{V}^\dagger-\alpha^* \hat{V}}\,,\qquad \alpha\in\mathbb{C}\,, \label{DispOp}
\end{equation}
on the most symmetrical state, which is the state $|0\rangle$ since $\hat{n}|0\rangle =0$. Writing $\alpha=re^{i\theta}$, it can be checked that $\hat D(\alpha)=\hat{U}^\theta(r)$.

The Displacement operator verifies the group homomorphism property:
\begin{equation}
  \hat D(\alpha)\hat D(\beta) = \hat D(\alpha+\beta)\,,\qquad \hat D(\alpha)^\dagger=D(-\alpha)\,.
\end{equation}

Then, the $E(2)$ coherent states are defined as:
\begin{eqnarray}
 |\alpha\rangle&=&\hat D(\alpha)|0\rangle =e^{\alpha \hat{V}^\dagger-\alpha^* \hat{V}}|0\rangle
 = e^{r \left(e^{i\theta}\hat{V}^\dagger- \left( e^{i\theta}\hat{V}^\dagger \right)^{-1}\right)}|0\rangle \nonumber \\
 &=& \sum_{n=-\infty}^{\infty} J_n(2r)   e^{in\theta} \hat{V}^{\dagger n} |0\rangle \label{CSE(2)}\\
 &=& \sum_{n=-\infty}^{\infty}
 \alpha^{n}2^{n}k_{n}(2|\alpha|)|n\rangle
 \equiv  \sum_{n=-\infty}^{\infty} c_n(r,\theta) |n\rangle \nonumber\,,
\end{eqnarray}
%
where\footnote{The functions $\frac{J_n(x)}{x^n}$ are known as Bochner-Riesz integral kernels \cite{Bochner-Riesz13}.} $k_n(x)=\frac{J_n(x)}{x^n}$, and
the generating function of Bessel functions have been used \cite{Gradshteyn}.
The coherent states can be written as $|\alpha\rangle =\sum_{n=0}^\infty \alpha^{n} h_n(|\alpha|^2)|n\rangle$ , and therefore  are coherent states of the AN type \cite{GazeauQO19},  with
$h_n(x)= 2^{n}k_{n}(2 \sqrt{x})$.

The radius of convergence of the series in  (\ref{CSE(2)}) is infinite, as can be checked 
using the asymptotic behaviour of Bessel functions for large $n$:
\begin{equation}
 J_n(x)\approx \frac{1}{\sqrt{2\pi n}} \left(\frac{e x}{2n}\right)^n \,. \label{asymptoticBessel}
\end{equation}
Therefore, this family of $E(2)$ coherent states is defined on the whole complex plane.

The coherent state overlap is:
\begin{equation}
 \langle\alpha|\alpha'\rangle = \langle re^{i\theta}|r'e^{i\theta'}\rangle
 = \sum_{n=-\infty}^{\infty} J_n(2r)J_n(2r')e^{in(\theta'-\theta)} = J_0(2 R)\,,\label{overlap}
\end{equation}
where the \textit{summation theorem} for Bessel functions (see \cite{Gradshteyn}) has been used, and where 
$R=\sqrt{r^2+{r'}^2-2rr'\cos(\theta'-\theta)}$.
Some specific values are:
\begin{eqnarray}
\langle r e^{i\theta}|r' e^{i\theta} \rangle & = &  J_0(2|r'-r|)  \\
 \langle r e^{i\theta}|r'e^{i(\theta+\frac{\pi}{2})}\rangle &=&  J_0(2\sqrt{r^2+r'{}^2}) \\
 \langle r e^{i\theta}|r' e^{i(\theta+\pi)} \rangle & = & J_0(2(r+r'))  \\
\langle r e^{i\theta}|r'e^{i(\theta+\frac{3\pi}{2})}\rangle &=& J_0(2\sqrt{r^2+r'{}^2}) \\
\langle r e^{i\theta}|r e^{i\theta'} \rangle & = & J_0(4r\sin\left(\frac{\theta'-\theta}{2}\right))  \\
\langle r e^{i\theta}|r e^{i\theta} \rangle & = & J_0(0)=1\,. \label{normalizationE2CS}
\end{eqnarray}
Note that by Eq.(\ref{normalizationE2CS}), these states are normalized.

Using the results of Sec. \ref{HWGA}, we can conclude that the  E(2) coherent state $|re^{i\theta}\rangle$ corresponds to the propagation (with $r$ playing the role of $z$), in a waveguide array with tilting angle $\theta$, of a Kronecker delta input. Although the intensity profiles for these states, corresponding to the usual diffraction pattern similar to that of propagation in empty space, are independent ot $\theta$ (as shown in  the left-most graphs in Figs. \ref{Figtheta0}, \ref{FigthetaPid4} and \ref{FigthetaPid2} in Sec. \ref{PhysicalInterpretation}), the angle $\theta$ will be relevant in the superposition of different coherent states, where relative phases are important, and the intensity profiles deviate notably from the usual difraction pattern.

\section{Resolution of the identity}
\label{ResolutionIdentity}

Once coherent states have been defined, we need to provide a resolution of the identity operator.  We shall recover the well-known result that the usual construction fails, the reason being that the group E(2) does not possess square integrable representations (see for instance \cite{Isham91}). Let us see it in detail.

\subsection{Naive resolution of the Identity}
\label{NaiveRI}

Coherent states of E(2) form a total (or overcomplete)  family in the Hilbert space $\cal H$. In the setting of waveguide arrays, any  light distribution in the infinite array (with a finite amount of energy for fixed $z$)  can be expressed in terms of coherent states $|\alpha\rangle, \alpha\in \mathbb{C}$. In addition, the coherent states
$|z e^{i\theta}\rangle$ represent themselves the  propagation in $z$  of light incident at    $z=0$ at the waveguide $n=0$ under the evolution of the tilted  Hamiltonian $\hat H_\theta$ given in Eq. (\ref{Htilted}).

Following the usual construction of a resolution of the identity for coherent states, and taking into account that the invariant measure under $E(2)$ on the complex plane is $d\alpha d\alpha^* = rdrd\theta$, we might write:
\begin{eqnarray}
 \hat{A} &=& \frac{1}{2\pi}\int_0^{2\pi}d\theta \int_0^\infty rdr |re^{i\theta}\rangle\langle re^{i\theta}| = \int_0^\infty rdr \sum_{n=-\infty}^\infty J_n(2r)^2 | n\rangle \langle n|\label{naiveResolution} \,.
\end{eqnarray}

But each of the integrals $\int_0^\infty rdr  J_n(2r)^2$ is divergent, a symptom of a high degree of redundancy of the family of coherent states. We can try with the measure $dr d\theta$, but this leads again to divergent integrals. With measures $d\theta \frac{dr}{r^\nu}$, with $\nu>0$, the integrals converge (see \cite{Gradshteyn}, formula 6.574-2), but the result depends on $n$, i.e.:
\begin{equation}
 \hat{A}_\nu=  \int_0^\infty \frac{dr}{r^\nu} \sum_{n=-\infty}^\infty J_n(2r)^2 | n\rangle \langle n| =  
 \frac{\Gamma(\nu)}{2\Gamma(\frac{\nu+1}{2})^2}\sum_{n=-\infty}^\infty \frac{\Gamma(|n|+\frac{1-\nu}{2})}{\Gamma(|n|+\frac{1+\nu}{2})}| n\rangle \langle n|\,.
\end{equation}

Thus $\hat{A}_\nu$ is not a resolution of the identity, and since its eigenvalues behave as $|n|^{-\nu}$ for large $|n|$,  it is not even a frame operator since the eigenvalues approach zero and therefore the inverse operator 
$\hat{A}_\nu^{-1}$, needed for the reconstruction of states, is unbounded (see the Appendix in \cite{Guerrero21b}). In addition, since the measure $d\theta \frac{dr}{r^\nu}$ is not invariant under $E(2)$, the important property of covariance of Perelomov CS is lost.

Note that 
\begin{equation}
 \lim_{\nu\rightarrow 0} \frac{2\Gamma(\frac{\nu+1}{2})^2}{\Gamma(\nu)} \hat{A}_\nu =\hat{I}_{\cal H}\,.
\end{equation}
However, this limit procedure can cause problems in explicit numerical computations.

Therefore, it is difficult to find an integration measure of this form. Following \cite{SphereMomentum20,CSE2}, we can introduce a scalar product for the Coherent State representation and a resolution of the identity based on the fact that $E(2)$ coherent states are solutions to the Helmholtz equation.

\subsection{Differential realization and the Helmholtz Equation}
\label{Helmholtzeq}

A differential realization for the $E(2)$ generators acting on the label space $\alpha=re^{i\theta}$  of the coherent states can be obtained in the usual way (see, for instance \cite{Miller68}), resulting in:
\begin{eqnarray}
 \hat{n}_d&=& -i \frac{\partial\,}{\partial\theta} \label{diff_n} \nonumber\\
 \hat{V}_d &=& -\frac{1}{2}e^{i\theta}\left(\frac{\partial\,}{\partial r} + \frac{i}{r}\frac{\partial\,}{\partial\theta}\right) \\
 \hat{V}_d^\dagger &=& \frac{1}{2}e^{-i\theta}\left(\frac{\partial\,}{\partial r} -\frac{i}{r}\frac{\partial\,}{\partial\theta}\right) \,.\nonumber
\end{eqnarray}

Note that the action of these operators on the coefficients of the Coherent States is:
\begin{eqnarray}
 \hat{n}_d\,c_n &=& n c_n\nonumber\\
 \hat{V}_d\, c_n &=& c_{n+1} \\
 \hat{V}_d^\dagger\, c_n &=& c_{n-1}\,,\nonumber
\end{eqnarray}
thus the role of lowering and raising operators are interchanged when acting on the coefficients instead of the ket vectors (as it should be, since mathematically this corresponds to taking the adjoint). However, Hermitian operators like the Hamiltonian are not affected by this interchange of roles.

We can also obtain the differential realization of the operators in Sec. \ref{HWGA}:
\begin{eqnarray}
 \hat{H}_d &=& \hat{V}_d^\dagger+\hat{V}_d= -i\frac{\partial\,}{\partial y} \nonumber\\
 \hat{H}_\theta{}_d &=& -i (e^{i\theta} \hat{V}_d^\dagger-e^{-i\theta}\hat{V}_d) = \frac{\partial\,}{\partial r} \,,
\end{eqnarray}
with $\alpha=re^{i\theta}=x+i y$.


From the differential realization, we derive that the eigenvalue equation for the quadratic Casimir of the Euclidean algebra, $\hat{C}_ 2=\hat{V}_d\hat{V}_d^\dagger=\hat{V}_d^\dagger\hat{V}_d=\hat I$, leads to the Helmholtz equation in the plane in polar coordinates:
\begin{equation}
 \hat{C}_ 2|\alpha\rangle =|\alpha\rangle\qquad \Rightarrow \qquad
\left( \frac{\partial^2\,}{\partial r^2} + \frac{1}{r} \frac{\partial\,}{\partial r}+
  \frac{1}{r^2} \frac{\partial^2\,}{\partial \theta^2} +k^2\right)|\alpha\rangle=0 \,,\label{HelmholtzPolar}
\end{equation}
where the wave number is $k=2$. Note also that each of the coefficients $c_n(r,\theta)$ satisfy the Helmholtz equation, constituting a basis for the Hilbert space ${\cal H}_{\rm osc}$ of bounded oscillatory solutions to the Helmholtz equation \cite{Wolf81,SphereMomentum20}, which is isomorphic to $\cal H$.

It should be stressed that only regular solutions at the origin appear in the coherent states (and thus on $c_n(r,\theta)$), consequently Bessel functions of  the second kind $Y_n$, which are also solutions to the Helmholtz equation (but unbounded), are discarded. This property is fundamental for the isomorphism between  $\cal H$ and  $\cal H_{\rm osc}$.

Seeing the complex number $\alpha=re^{i\theta}=x+i\:y$ as a vector $\vec{\alpha}=(x,y)\in\mathbb{R}^2$, Eq. (\ref{HelmholtzPolar}) is written as
\begin{equation}
 \left[\Delta_{\vec{\alpha}}+k^2 \right]|\alpha\rangle=0\,. \label{HelmholtzCartesian}
\end{equation}

\subsection{Resolution of the identity in terms of Cauchy coherent states}
\label{resolutioncartesian}

The Helmholtz equation is an elliptic PDE and, as such, the Cauchy problem for it ill-posed \cite{Payne}. However, a well-posed Cauchy problem can be defined if we restrict the space of solutions to that of oscillatory solutions \cite{Wolf81}. According to this, since coherent states are oscillatory solutions to the Helmholtz equation, Eq. (\ref{HelmholtzPolar})  or (\ref{HelmholtzCartesian}) implies that any coherent state can be obtained from coherent states with labels restricted to a line  (together with their normal derivatives, seen as  a Cauchy initial value problem, see \cite{Wolf81} and Appendix B of  \cite{SphereMomentum20}).
This redundancy explains why integration on the whole complex plane (see Eq. (\ref{naiveResolution})) causes divergences. In fact, the normal direction to the Cauchy line represents the propagation of the oscillatory solution, and integrating on the propagation variable for an oscillatory solution to the Helmholtz equation always produces a divergence\footnote{This is a general fact, that applies also to parabolic equations like the Schr\"odinger equation and hyperbolic equations like the Klein-Gordon equation, where the integration is always with respect to position or momentum variables, but never with respect to time, since unitary evolution implies that the integral is infinite. It also applies to the propagation of light in the waveguide array, as it was noted at the end of Sec. \ref{HWGA}.}.
We can try with different integration measures, but the problem is not the measure, but the redundancy.

Although it is possible to characterize solutions to the Helmholtz equation by their values on some boundary region, in this paper  we shall restrict to the case of Cauchy initial conditions, leaving the more complex case of boundary conditions to a future work.

In fact, any oscillatory solution to the Helmholtz equation (\ref{HelmholtzCartesian}) can be expressed\footnote{See \cite{SphereMomentum20}, Eq. (B.10)-(B.14) in Appendix D, for a derivation of an equivalent expression in Fourier (momentum) space.}
in terms of its values and those of  its derivative with respect to, let us say, $y$ at $y=0$:
\begin{eqnarray}
 \psi(x,y)&=&\int_\mathbb{R} dx'\left[ \Delta(x-x',y-y')\frac{\partial\psi(x',y')}{\partial y'}-\frac{\partial \Delta(x-x',y-y')}{\partial y'}\psi(x',y')\right]_{y'=0} \nonumber \\
  &=& \int_\mathbb{R} dx'\left(\Delta(x-x',y) \psib(x') + \frac{\partial \Delta(x-x',y)}{\partial y} \psic(x')\right)\,, \label{PropagationHelmholtzRectangular}
\end{eqnarray}
where 
\begin{eqnarray}
 \Delta(x,y)&=&\frac{1}{\sqrt{2\pi}}\int_{-k}^k d\epsilon  e^{i \epsilon x} \frac{\sin(\sqrt{k^2-\epsilon^2}y)}{\sqrt{k^2-\epsilon^2}} \nonumber \\
 &=&
 \sqrt{\frac{2}{\pi}}k \frac{\sin(y\sqrt{k^2+\frac{\partial^2\,}{\partial x^2} })}{\sqrt{k^2+\frac{\partial^2\,}{\partial x^2}}}  j_0(k x) \label{PropagatorHelmholtzCartesian} \\
 &=& \sqrt{\frac{2}{\pi}}k y \sum_{n=0}^\infty  (-1)^n \frac{n!}{(2n+1)!} \left( \frac{2ky^2}{x}\right)^n j_n(kx) \nonumber
\end{eqnarray}
is the Helmholtz (self-)propagator (i.e. a fundamental solution to the Helmholtz equation without sources), and $\psic(x)$  and $\psib(x)$ are the values of $\psi(x,y)$ and its derivative with respect to $y$ at $y=0$, respectively. In the last formula $j_n$ is the spherical Bessel function, which for $n=0$ coincides with the \textit{sinc} function $\frac{\sin x}{x}$. Using (\ref{asymptoticBessel}), it is easy to check that the series is absolutely (and uniformly) convergent for all values of $x$ and $y$, therefore the Helmholtz propagator is well defined.

It should be remarked that, being  $\Delta(x,y)$ a solution to the Helmholtz equation, it satisfies Eq. (\ref{PropagationHelmholtzRectangular}) with Cauchy initial data $\Deltac(x)=0$ and 
$\Deltab(x)=\sqrt{\frac{2}{\pi}}k j_0(k x) $.


Equation (\ref{PropagationHelmholtzRectangular}) implies a high redundancy for $\psi(x,y)$ (in particular for coherent states, which are also solutions to the Helmholtz equation), thus it is clear that any integration involving $\psi(x,y)$ with the invariant measure under $E(2)$, $dxdy$, will be divergent.

To avoid this redundancy, we follow the ideas in \cite{Wolf81} (see also \cite{SphereMomentum20,CSE2}), where a scalar product is introduced involving only the values of the functions and their derivatives with respect to $y$ at $y=0$, and with a double integral with a convolution kernel. For this purpose,  let us introduce two subsets of states:
\begin{eqnarray}\label{CSE2cartesian}
 |\!\xcirc\: \rangle &=& |x+i\:y\rangle|_{y=0}= \sum_{n=-\infty}^{\infty} J_n(2x) |n\rangle \nonumber\\
 |\!\xbullet\: \rangle &=& -\frac{i}{\sqrt{2}}\frac{\partial\,}{\partial y}|x+i\:y\rangle|_{y=0} = \frac{1}{\sqrt{2}} \hat H_d|x+i\:y\rangle|_{y=0}= \frac{1}{\sqrt{2}}\sum_{n=-\infty}^{\infty} n \frac{J_n(2x)}{x} |n\rangle \,.
 \label{initialDataCartesian}
\end{eqnarray}

The states $|\!\xcirc\: \rangle$ are a particular class of coherent states, but $|\!\xbullet\: \rangle$ are not coherent states for any value of $x$ (see the end of the section for an interpretation of these states). However, the pair $(|\!\xcirc\: \rangle,|\!\xbullet\: \rangle)$ corresponds to the Cauchy initial data needed in Eq. (\ref{PropagationHelmholtzRectangular}) to recover an arbitrary coherent state, thus any coherent state can be expressed in terms of them, constituting a generating system:
\begin{equation}
 |x+i y\rangle 
  = \int_\mathbb{R} dx'\left(\Delta(x-x',y) |\!\xbullet{}'\: \rangle + \frac{\partial \Delta(x-x',y)}{\partial y}|\!\xcirc{}'\: \rangle \right)\,. \label{PropagationCSRectangular}
\end{equation}

For this reason, we shall denote the pair $(|\!\xcirc\: \rangle,|\!\xbullet\: \rangle)$ as  \textit{Cauchy coherent states} for the Euclidean group $E(2)$.

The overlaps among these states are:
\begin{eqnarray} 
 \langle \xcirc|\!\xcirc{}' \rangle & = &  J_0(2|x'-x|) =  k_0(2|x'-x|) \nonumber \\
 \langle \xbullet |\!\xbullet{}'\rangle &=&  \frac{1}{2xx'} \sum_{n=-\infty}^{\infty} n^2    J_n(2x)J_n(2x') = \frac{J_1(2|x'-x|)}{|x'-x|} = 2 k_1(2|x'-x|) \label{overlapcartesian}\\
 \langle \xcirc|\!\xbullet{}' \rangle & = & 0 \,,  \nonumber 
\end{eqnarray}
from which we deduce that $|\!\xcirc \:\rangle$ and $|\!\xbullet \:\rangle$ are normalized and orthogonal\footnote{The fact that $|\!\xcirc \:\rangle$ and $|\!\xbullet \:\rangle$ are orthogonal implies that both sets of states are required to expand the whole Hilbert space ${\cal H}$.} with respect to each other.
Denoting $\Hcirc={\rm span}\{|\!\xcirc \:\rangle,\,x\in\mathbb{R}\}$ and $\Hbullet={\rm span}\{|\!\xbullet \:\rangle,\,x\in\mathbb{R}\}$, according to the previous discussion we have that ${\cal H}=\Hcirc \oplus \Hbullet$.

The resolution operator for Cauchy CS is given by:
\begin{equation}
 \hat{A}_{\rm C} = \int_\mathbb{R} dx \left(|\!\xcirc \:\rangle \langle \xcirc|+|\!\xbullet\rangle \langle \xbullet|\right)
 = \Acirc_{\rm C} \oplus \Abullet_{\rm C} \,,
\end{equation}
where $\Acirc_{\rm C}$ and $\Abullet_{\rm C}$ are the restrictions of the resolution operator to the subspaces $\Hcirc$  and $\Hbullet$, respectively. For the integration measure we have chosen the Euclidean (E(1)) invariant measure $dx$.

Both $\Acirc_{\rm C}$ and $\Abullet_{\rm C}$, by construction, are integral (convolution, in fact) operators with kernels $k(x,x')$ given by $k_0(2|x'-x|)$ and $2 k_1(2|x'-x|)$, respectively. Let us denote by $\hat{K}_\alpha$ the convolution operator with kernel $k_\alpha(2|x'-x|)$. Then
\begin{equation}
 \Acirc_{\rm C} = \hat{K}_0|_{\Hcirc}\,,\qquad \Abullet_{\rm C}=2 \hat{K}_1|_{\Hbullet} \,.
\end{equation}

In addition, since their convolution kernels are Bochner-Riesz integral kernels \cite{Bochner-Riesz13}, the operators $\Acirc_{\rm C}$ and $\Abullet_{\rm C}$ are positive definite (despite the fact that the kernels are oscillatory functions), bounded and invertible. We can obtain a resolution of the identity operator using the theory of frames (see the Appendix in
\cite{Guerrero21b} and references therein), using the concept of dual frame, and for that purpose the inverse of the resolution operator is required.

The inverse opertators are also integral operators with convolution kernels given by the inverse (under convolution) of the corresponding kernels, which can be easily computed using the formula:
\begin{equation}
 \int_\mathbb{R}dx'' k_\alpha (|x-x''|)k_\beta(|x''-x'|)=\frac{N_{\alpha-1/2}N_{\beta-1/2}}{N_{\alpha+\beta-1}}k_{\alpha+\beta-1/2}(|x-x'|)\,,
 \label{convolution2kernels}
\end{equation}
with $N_{\alpha}=\frac{1}{2^\alpha\Gamma(\alpha+1)}$ (see Appendix D of Ref. \cite{SphereMomentum20}), and taking into the account that the identity operator in ${\cal H}$ is the convolution operator $2 \hat{K}_\frac{1}{2}$ (acting on both subspaces $\Hcirc$ and $\Hbullet$), with kernel $2 k_\frac{1}{2}(2|x'-x|)$ and which is simply the projector onto ${\cal H}$ (see Appendix C of Ref. \cite{SphereMomentum20}).

The inverses $\Acirc_{\rm C}{}^{-1}$ and $\Abullet_{\rm C}{}^{-1}$ are then:
\begin{equation}
\Acirc_{\rm C}{}^{-1} = \pi\, 2\hat{K}_1|_{\Hcirc}\,,\qquad \Abullet_{\rm C}{}^{-1}= \pi \hat{K}_0|_{\Hbullet}\,.
\end{equation}

Although the resolution operator is not the identity operator in ${\cal H}$, the fact that it is bounded, invertible and with a bounded inverse, allows to define a dual family of coherent states (the dual frame, see the Appendix in \cite{Guerrero21b}). These are defined by:
\begin{eqnarray}
 \widetilde{|\!\xcirc\: \rangle } &=&     \pi\, 2\hat{K}_1|\!\xcirc\: \rangle  \\
 \widetilde{|\!\xbullet\: \rangle } &=&\pi \hat{K}_0|\!\xbullet\: \rangle\,.
\end{eqnarray}

With the dual frame, we can build a resolution of the Projector:
\begin{eqnarray}
 \hat{P}  &=&\int_\mathbb{R} dx \left( \widetilde{|\!\xcirc \:\rangle} \langle \xcirc| + \widetilde{|\!\xbullet \:\rangle} \langle \xbullet| \right)\nonumber \\
 &=&  \pi \int_\mathbb{R} dx \int_\mathbb{R} dx'  \left(2 k_1(2|x'-x|) |\!\xcirc \:\rangle \langle \xcirc{}'| + k_0(2|x'-x|) |\!\xbullet\rangle \langle \xbullet{}'|\right)  =2  \hat{K}_\frac{1}{2}\,. \label{ResolutionE2cartesian}
\end{eqnarray}
which coincides with the Identity ${I}_{\cal H}$ on ${\cal H}$.


We can check the validity of this statement  by considering the matrix elements in the $|\!\xcirc \:\rangle$ and $|\!\xbullet \:\rangle$ basis:
\begin{eqnarray}
 \langle \xcirc|\hat P|\!\xcirc{}' \rangle& =& k_0(2|x-x'|)\nonumber\\
  \langle \xbullet|\hat P|\!\xbullet{}' \rangle& =& 2 k_1(2|x-x'|)\\
  \langle \xcirc|\hat P|\!\xbullet{}' \rangle& =& 0  \,,\nonumber
\end{eqnarray}
which implies, since the Cauchy coherent states (\ref{initialDataCartesian}) are overcomplete, and according to (\ref{overlapcartesian}),
that $\hat P={I}_{\cal H}$, and where we have used eqn. (\ref{convolution2kernels}).


Due to the Euclidean symmetry E(2) of the non-local integration appearing in the resolution of the identity (\ref{ResolutionE2cartesian}) \cite{Wolf81,SphereMomentum20}, the construction of coherent states should be valid using as initial Cauchy set any line obtained by translation and/or rotation of the line $y=0$ used in Eqs. (\ref{CSE2cartesian}). Thus, we can also use the families of coherent states:
\begin{eqnarray} \label{CSE2rotated}
|\!\ucirc\: \rangle_{\theta_0}&=& |u e^{i\theta}\rangle|_{\theta=\theta_0}= \sum_{n=-\infty}^{\infty} e^{i n \theta_0}J_n(2u) |n\rangle \nonumber\\
 |\! \ubullet\: \rangle_{\theta_0}&=& -\frac{i}{\sqrt{2}u}\frac{\partial\,}{\partial \theta}|u e^{i \theta}\rangle|_{\theta=\theta_0} = \frac{1}{\sqrt{2}} (e^{i\theta}\hat V_d^\dagger+e^{-i\theta}\hat V_d) |u e^{i\theta}\rangle|_{\theta=\theta_0} \\
 &=& \frac{1}{\sqrt{2}}\sum_{n=-\infty}^{\infty} n e^{i n \theta_0}\frac{J_n(2u)}{u} |n\rangle \nonumber
\end{eqnarray}
with\footnote{We requiere $u=\pm r$ and restrict $\theta_0$ to the interval $[0,\pi)$ to pass from polar coordinates to  rotated cartesian coordinates.} $u=\pm r= r e^{i\frac{1\mp 1}{2}\pi}$ and $\theta_0\in [0,\pi)$. Then equations (\ref{overlapcartesian}) and (\ref{ResolutionE2cartesian}) are valid substituting the coherent states in (\ref{CSE2cartesian}) by the ones in (\ref{CSE2rotated}).

In particular, for $\theta_0=\frac{\pi}{2}$, we recover that the coherent states  $|\!\ucirc\: \rangle_{\frac{\pi}{2}}$ are the states obtained by the (forward/backward) propagation  of light impinged at $x=0$ (i.e. the coordinate $x$ plays the role of the propagation coordinate $z$ in the waveguide array) at the waveguide $n=0$ under the Hamiltonian $\hat H$. This is a remarkable fact: these coherent states have a natural physical realization in a waveguide array (see \cite{Pertsch99} and also \cite{RodriguezLara18}). In addition, in order to obtain a resolution of the Identity in $\cal H$, we also need the states $|\! \ubullet\: \rangle_{\frac{\pi}{2}}$, which are obtained by acting with the operator $\frac{i}{\sqrt{2}} (\hat V_d^\dagger-\hat V_d)=\frac{i}{\sqrt{2}} \frac{\partial\ }{\partial x}$ on the coherent states and restricting to $x=0$. For other values of $\theta$, the states $|\!\ucirc\: \rangle_{\theta}$ are realized in tilted waveguide arrays.

An important comment is in order. The reason why we need two families of states to obtain a resolution of the Identity, $(|\!\xcirc\: \rangle,|\!\xbullet\: \rangle)$ (or $(|\!\ucirc\: \rangle_{\theta_0}, |\! \ubullet\: \rangle_{\theta_0})$), one of them corresponding to the restriction of coherent states to a line, while the other is obtained by the action of some operator on them, can be traced back to the  fact that coherent states satisfy the Helmholtz equation and we are using initial conditions $(\psic(x),\psib(x))$ to propagate any oscillatory solution $\psi(x,y)$ by means of (\ref{PropagationHelmholtzRectangular}), since the Helmholtz equation is a second order elliptic PDE.


In Sec. \ref{PhysicalInterpretation} we provide a physical interpretation of Cauchy coherent states in tilted waveguide arrays.

\subsection{Polar Cauchy coherent states}

Although we shall not discuss them in detail in this paper, leaving it for a future publication, we shall introduce Cauchy coherent states when the Cauchy set is a circunference centered at the origin.

One of the most common ways of solving the Helmholtz equation is with boundary conditions in a circumference centered at the origin. Depending on whether we are interested in obaining the solution in the inner or the outer region of the circunference, we obtain Bessel functions or Hankel functions (if Sommerfeld radiation condition is imposed), respectively. However, we are interested here in the case of Cauchy initial conditions in the circumference.

Let us introduce the analogous notion of Cauchy coherent states in this case:

\begin{eqnarray}
 |\!\!\thetacirc\rangle_{r_0} &=& |r_0e^{i\theta}\rangle\nonumber  \\
 |\!\!\thetabullet\rangle_{r_0} &=& \sqrt{2} \left(\frac{\partial\,}{\partial r} |re^{i\theta}\rangle\right)_{r=r_0} = \sqrt{2} \sum_{n=-\infty}^{\infty}  e^{i n \theta } J_n'(2 r_0)|n\rangle
 \label{PolarCauchyCS}
\end{eqnarray}
where $r_0$ is the radius of the circumference. The scalar product among these states are:
\begin{eqnarray}
 {}_{r_0}\langle\thetacirc\!|\!\thetacirc{\!}'\rangle_{r_0} &=&  \sum_{n=-\infty}^{\infty} J_n(2r_0)^2 e^{in(\theta'-\theta)} = k_0(4r_0 \sin (\frac{\theta'-\theta}{2})) \\
{}_{r_0}\langle\thetabullet\!|\!\thetabullet{\!}'\rangle_{r_0} &=& 2\sum_{n=-\infty}^{\infty} J_n'(2r_0)^2 e^{in(\theta'-\theta)} \nonumber \\
 &=&  2 k_1(4r_0 \sin (\frac{\theta'-\theta}{2}))- 2 k_0(4r_0 \sin (\frac{\theta'-\theta}{2}))\sin^2 (\frac{\theta'-\theta}{2}) \,.
\end{eqnarray}

Note that this states are normalized. However, they are  not orthogonal,

\begin{equation}
 {}_{r_0}\langle\thetacirc\!|\!\thetabullet{\!}'\rangle_{r_0} = -4 r_0 k_1(4r_0 \sin (\frac{\theta'-\theta}{2}))\sin^2 (\frac{\theta'-\theta}{2})\,,
\end{equation}
except for $\theta=\theta'$.


Let us compare polar Cauchy coherent states (\ref{PolarCauchyCS}) with (rotated) cartesian Cauchy coherent states (\ref{CSE2rotated}). To simplify the notation, we shall take $\theta_0=\theta$ and $r_0=r$, and consider $u=r$. With this notation, we have:
\begin{eqnarray}
  |\!\thetacirc\rangle_{r} &=& |\!\rcirc\rangle_{\theta} = |re^{i\theta}\rangle \\
   {}_{r}\langle\thetabullet\!|\!\rbullet{}'\rangle_{\theta'} &=& -\frac{2\sqrt{2}\sin(\theta'-\theta)}{R^2}\left[(r^2-r'{}^2)k_1(2R)-r k_0(2R)(r-r'\cos(\theta'-\theta))\right]
\end{eqnarray}
where $R$ is defined after Eq. (\ref{overlap}). Note that when $\theta'-\theta=0,\pi$ the states $|\!\rbullet{}'\rangle_{\theta'}$ and $|\!\thetabullet\rangle_{r}$ are orthogonal, reflecting the fact that
the tangent vector $\frac{\partial\,}{\partial r}$ at $(r,\theta)$ and $\frac{\partial\,}{\partial \theta}$ at $(r',\theta')$ are orthogonal only if $\theta'-\theta=0,\pi$. It is curiuos how the geometrical properties of the label space translate into the coherent states. 

\section{Physical interpretation of coherent states and Cauchy coherent states in tilted waveguide arrays}
\label{PhysicalInterpretation}


It is interesting to compare the propagation of light for different tilting angles $\theta$ and different (at $z=0$) light configurations among the waveguides 
given by a Gaussian profile with different widths. The Gaussian input state is given by:
\begin{equation}
 |\sigma\rangle =N_\sigma \sum_{n\in \mathbb{Z}} e^{-\frac{n^2}{2\sigma^2}} |n\rangle\,,
\end{equation}
where $N_\sigma$ is a normalization factor. The propagated state with the tilted Hamiltonian $\hat{H}^\theta$ is:
\begin{eqnarray}
 |\sigma,r e^{i\theta}\rangle &=& U^\theta(r) |\sigma\rangle \nonumber \\
 &=& N_\sigma \left\{ |r e^{i\theta}\rangle
 + \sum_{n\in \mathbb{Z}} e^{i n \theta} \sum_{m=1}^{\infty} e^{-\frac{m^2}{2\sigma^2}} \left( e^{-i m\theta} J_{n-m}(2r)+ e^{i m\theta} J_{n+m}(2r)\right) |n\rangle\right\} \,. \label{PropagationGaussian}
\end{eqnarray}

\begin{center}
\begin{figure}[h!]
\includegraphics[width=\graphwidth]{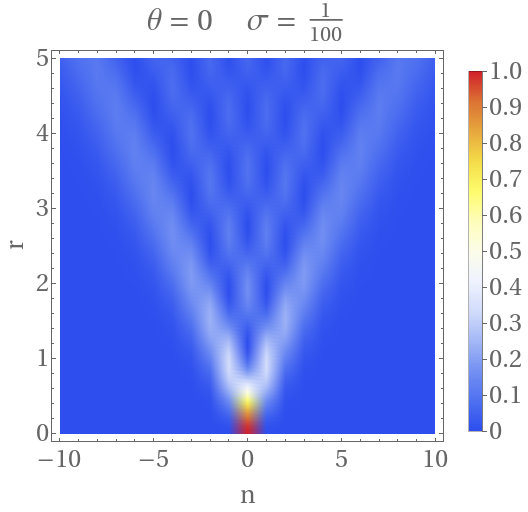}\hspace{\graphsep}\includegraphics[width=\graphwidth]{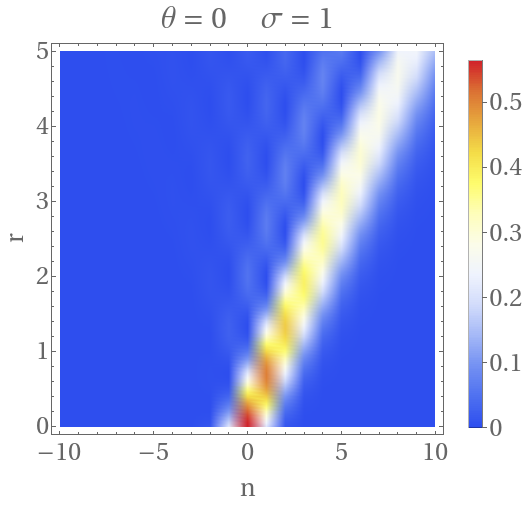}\hspace{\graphsep}
\includegraphics[width=\graphwidth]{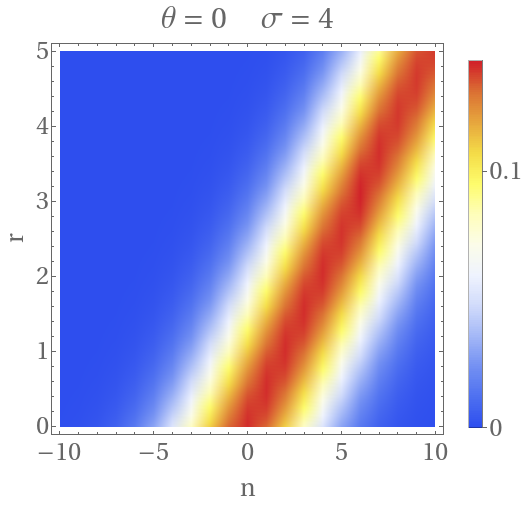}
\caption{Intensity distributions for $\theta=0$ and  $\sigma=\frac{1}{100}$ (left),  $\sigma=1$ (center) and $\sigma=4$  (right). Note that the colormap scale ranges from 0 to the maximum intensity  of the corresponding plot.}
\label{Figtheta0}
\end{figure}
\end{center}

\begin{center}
\begin{figure}[h!]
\includegraphics[width=\graphwidth]{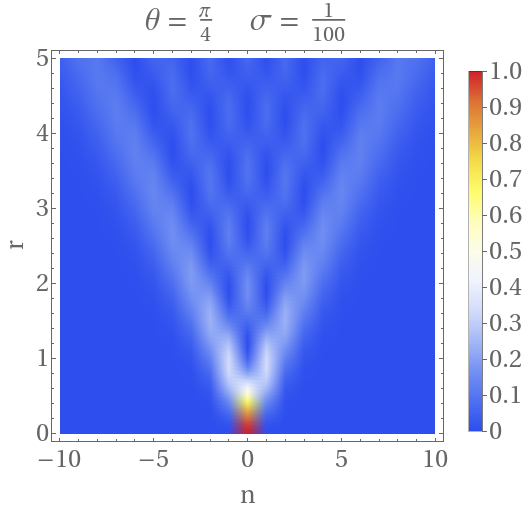}\hspace{\graphsep}\includegraphics[width=\graphwidth]{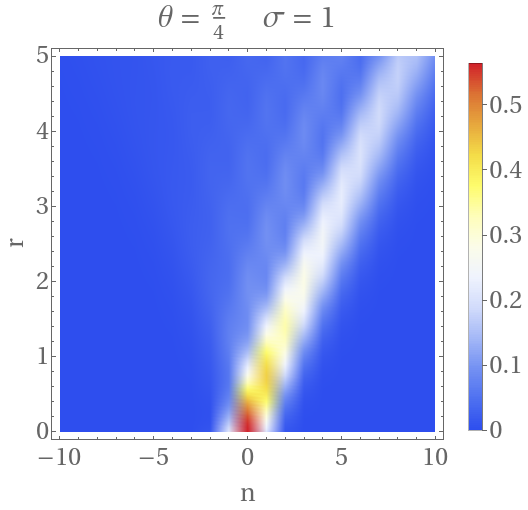}\hspace{\graphsep}
\includegraphics[width=\graphwidth]{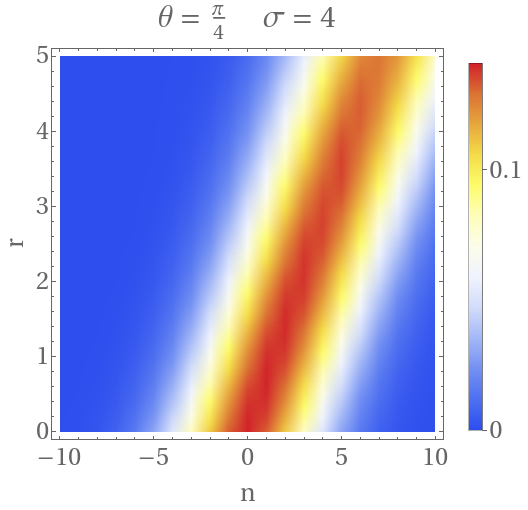}
\caption{Intensity distributions for $\theta=\frac{\pi}{4}$ and  $\sigma=\frac{1}{100}$ (left),  $\sigma=1$ (center) and  $\sigma=4$ (right). Note that the colormap scale ranges from 0 to the maximum intensity  of the corresponding plot.}
\label{FigthetaPid4}
\end{figure}
\end{center}

\begin{center}
\begin{figure}[h!]
\includegraphics[width=\graphwidth]{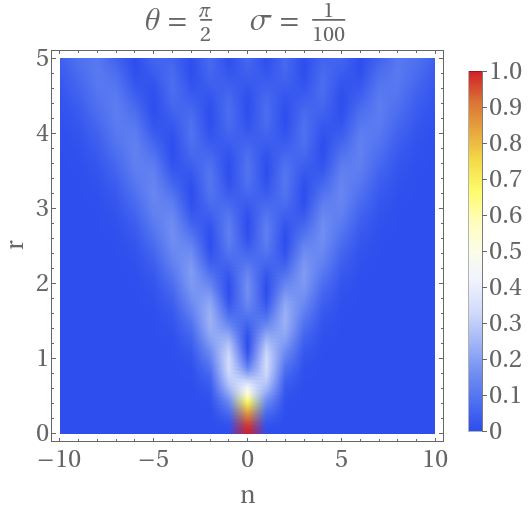}\hspace{\graphsep}\includegraphics[width=\graphwidth]{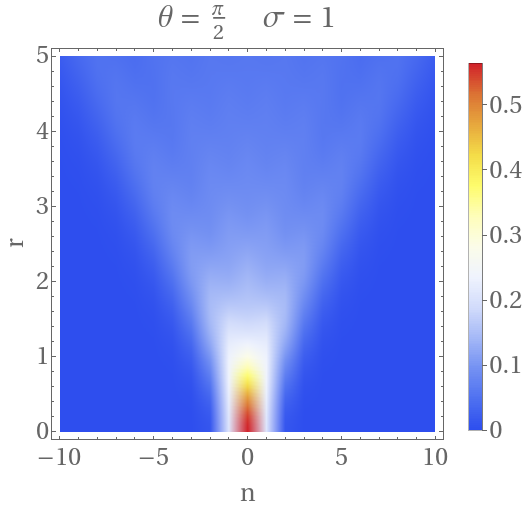}\hspace{\graphsep}
\includegraphics[width=\graphwidth]{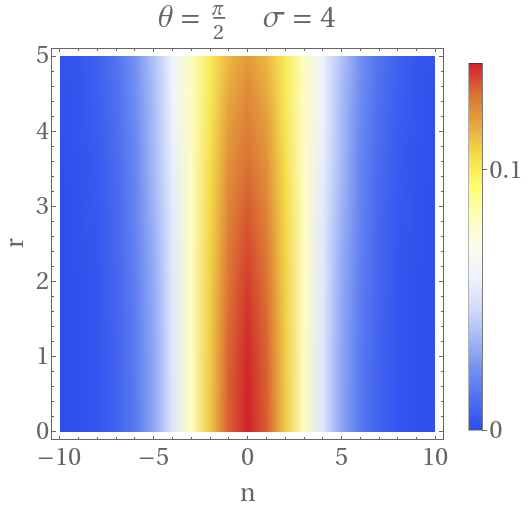}\
\caption{Intensity distributions for $\theta=\frac{\pi}{2}$ and  $\sigma=\frac{1}{100}$ (left), $\sigma=1$  (center) and  $\sigma=4$ (right). Note that the colormap scale ranges from 0 to the maximum intensity of the corresponding plot.}
\label{FigthetaPid2}
\end{figure}
\end{center}

In Figures \ref{Figtheta0}, \ref{FigthetaPid4} and \ref{FigthetaPid2}
the light intensity configurations in a (truncated) array of 21 waveguides with a length of 5 units (in units of inverse of the couplig constant) is shown for tilting angles $\theta=0$,  $\theta=\frac{\pi}{4}$ and  $\theta=\frac{\pi}{2}$,
and for  initial light configurations given by a Gaussian profile with different widths: $\sigma=1/100$ (essentially a Kronecker delta input at the $n=0$ waveguide), $\sigma=2$ (approx. 3 waveguides width) and $\sigma=4$ (approx. 7 waveguides width).

For tilting angles $\theta=\frac{3\pi}{4}$ and $\theta=\pi$,  the behaviour is simetrical with respect to the $n=0$ waveguide to that of the cases $\theta=\frac{\pi}{4}$ and  $\theta=0$, respectively.  Thus,  the behaviour is that of a cosine function with $2\pi$ periodicity.

While for the Kronecker delta input the intensity distribution is the same for all tilting angles $\theta$ (the reason being that
the light amplitudes are $A_n(z)=e^{i n \theta}J_n(2z)$, whose moduli do not depend on $\theta$), for larger values of $\sigma$ there are interference effects among the different waveguides resulting in tilted propagation (with respect to the $z$ axis). The maximun tilting is produced for $\theta=0$ and $\pi$ (corresponging to real tilting angles   $\varphi=\frac{3\pi}{2}$ (mod $2\pi$) and $\varphi=\frac{\pi}{2}$  see \cite{PhysRevLett.85.1863,PhysRevLett.88.093901}). For these values we also have zero diffraction for a sufficiently wide input profile. This property is known as \textit{anomalous diffraction} \cite{PhysRevLett.85.1863,PhysRevLett.88.093901}.

The states $|\! \rbullet\: \rangle_{\theta}$ and $|\!\!\thetabullet\rangle_{r}$ can be given an interpretation as the first-order correction to the approximation of the propagation of a Gaussian in a tilted waveguide array. To see this, in eqn. (\ref{PropagationGaussian}) we use  the fast decreasing property of the Gaussian to retain  only the terms with $m=0, \pm 1$ and 
use the approximation
\begin{equation}
 e^{-\frac{1}{2\sigma^2}} \approx 1-\frac{1}{2\sigma^2} \,.
\end{equation}
This approximation is rather good if $\sigma$ is near 1,  arriving at:
\begin{equation}
 |\sigma,r e^{i\theta}\rangle \approx 
 N_\sigma \left\{ |r e^{i\theta}\rangle
 + (1-\frac{1}{2\sigma^2})\sum_{n\in \mathbb{Z}} e^{i n \theta}  \left( e^{-i\theta} J_{n-1}(2r)+ e^{i \theta} J_{n+1}(2r)\right) |n\rangle\right\} \,.
\end{equation}

Finally, using the recurrence properties of Bessel functions we obtain:
\begin{eqnarray}
 |\sigma,r e^{i\theta}\rangle &\approx&
 N_\sigma \left\{ |r e^{i\theta}\rangle
 + (1-\frac{1}{2\sigma^2})\sum_{n\in \mathbb{Z}} e^{i n \theta}  \left( \frac{n}{r}J_{n}(2r)\cos\theta - i  J_{n}'(2r)\sin\theta \right)|n\rangle\right\} \nonumber\\
 & =& N_\sigma \left\{ |r e^{i\theta}\rangle
 + (1-\frac{1}{2\sigma^2}) \left( \sqrt{2}\cos\theta |\! \rbullet\: \rangle_{\theta} - i  \sin\theta |\!\!\thetabullet\rangle_{r}\right)\right\} \label{ApproximatedGaussian}
 \,.
\end{eqnarray}

Thus, we have proven that the evolution of an initial Gaussian state in a tilted waveguide array (with angle $\theta$) can be described at zero-th order (in $m$) by a coherent state $|r e^{i\theta}\rangle$ and at first order by a linear combination (rotation  by $\theta$ up to some numerical factors) of the states $|\! \rbullet\: \rangle_{\theta}$ and $|\!\!\thetabullet\rangle_{r}$. This simple formula contains the initial Cauchy data for both cartesian and polar coordinates, and provides a physical realization of them in tilted waveguide arrays.

\begin{center}
\begin{figure}[h!]
\includegraphics[width=\graphwidth]{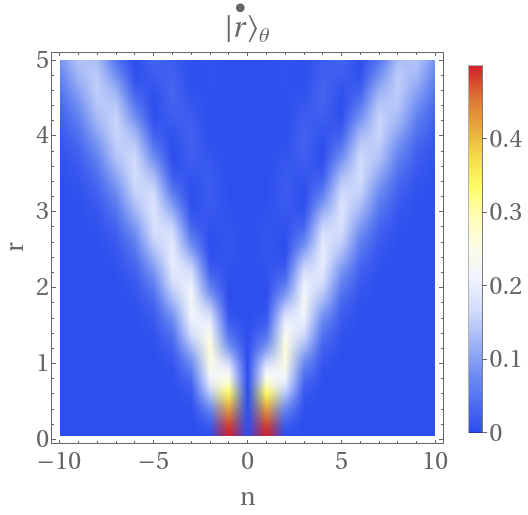}\hspace{\graphsep}
\includegraphics[width=\graphwidth]{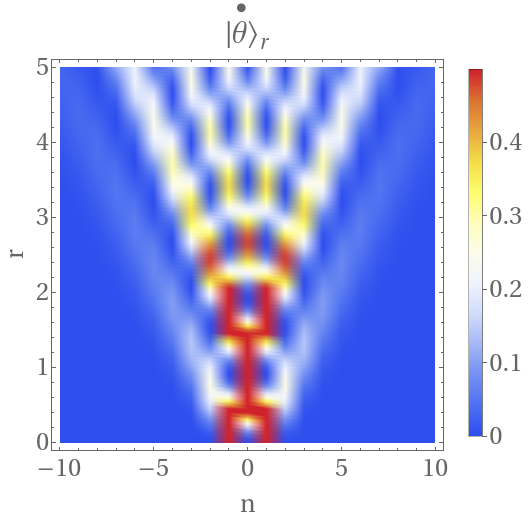}\hspace{\graphsep}
\includegraphics[width=\graphwidth]{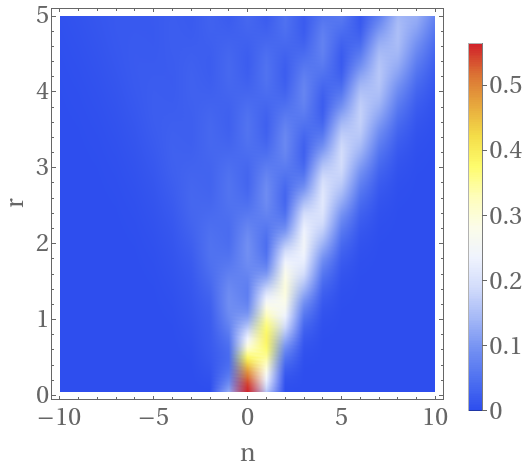}
\caption{Intensity distributions for $|\!\!\rbullet\rangle_{\theta}$ (left) and $|\!\!\thetabullet\rangle_{r}$ (center). Approximated propagation of a Gaussian input for $\sigma=1$ and $\theta=\frac{\pi}{4}$ given by eqn. (\ref{ApproximatedGaussian}) (right). Note that the colormap scale ranges from 0 to the maximum intensity of the corresponding plot.}
\label{CSbullet}
\end{figure}
\end{center}

In Figure \ref{CSbullet} (right), the intensity distribution of the approximated propagation of a Gaussian input for $\sigma=1$ and $\theta=\frac{\pi}{4}$ is shown, which should be compared with Figure \ref{FigthetaPid4} (center). The states $|\! \rbullet\: \rangle_{\theta}$ (left) and $|\!\!\thetabullet\rangle_{r}$ (center) are also shown.

Note that this construction can be generalized to any wave-packet initial state (non-necessarily Gaussian), with $\sigma$ being now the standard deviation (width) of the initial wave-packet.


\section{Conclusion}
\label{conclusions}

In this paper we have reviewed the construction of Perelemov coherent states of the Euclidean $E(2)$ group realized in an infinite array of homogeneous and equally spaced waveguide arrays.  The obtained  coherent states (\ref{CSE(2)}) are labeled only by a complex number $z$, or the pair $(x,y)$ with $z=x+i y$, that represents the momentum space of the E(2) group (see \cite{SphereMomentum20}). The angle variable $\varphi\in[0,2\pi)$ of the Euclidean group, representing configuration space, is absent, the reason being that we have chosen as \textit{fiducial state} the most \textit{symmetrical} one, i.e. the state $|0\rangle$ satisfying  $\hat{n}|0\rangle =0$, and $\hat{n}$ is precisely the generator of the compact subgroup of rotations\footnote{The reader should not confuse this fact with the expression of $\hat{n}$ in the differential realization $\hat{n}_d$
acting on functions on the label space variable $\alpha=r e^{i\theta}\in\mathbb{C}$. Due to the semidirect action of rotations on the Euclidean plane, $\hat{n}$ acts both on the compact variable $\varphi$ of E(2) and as rotations on the plane, (\ref{diff_n}), and this last one is the only surviving the condition $\hat{n}|0\rangle =0$.}
in $E(2)$. Choosing as fiducial state $|f\rangle =\sum_{n\in\mathbb{Z}} a_n |n\rangle$, with $\sum_{n\in\mathbb{Z}} |a_n|^2<\infty$ would require to add the angle $\varphi\in[0,2\pi)$ in the coherent states by adding the number operator $\hat{n}$ in the displacement operator $\hat{D}$ in eqn. (\ref{DispOp}), and the corresponding integration with respect to the angle $\varphi\in[0,2\pi)$ in the resolution of the identity given in eqn. (\ref{ResolutionE2cartesian}).

In  Sec. \ref{resolutioncartesian} we have provided a resolution of the identity  involving just coherent states on half the coherent state label space (at the price of including also the derivatives), i.e. in terms of the initial Cauchy data, eqn. (\ref{initialDataCartesian}), of the coherent states seen as  oscillatory solutions to the Helmholtz equation in label space (\ref{HelmholtzPolar}), or (\ref{HelmholtzCartesian}), and for this reason we named them \textit{Cauchy coherent states}. This construction is the same as the one provided in \cite{SphereMomentum20} (eq. (53)) for the momentum representation of a particle on the sphere (there on $S^3$, but this applies to any Euclidean group $E(n)$, in particular to $E(2)$). In fact, the states constructed there satisfy the same overlap (cf. Eq. (52) of \cite{SphereMomentum20}) as
in (\ref{overlapcartesian}), except for a shift in the index of the kernels due to the dimensionality of the sphere.

Due to this analogy, it would be interesting to construct the generalized Fourier transform (see \cite{SphereMomentum20}) connecting functions on the  label space $(x,y)$ (momentum space) with a representation in the circle (configuration space in terms of the variable of rotations in $E(2)$), and obtain a realization in terms of functions on the circle (and the corresponding coherent states) for solutions on the infinite waveguide array, providing an interpretation of it. This is work in progress \cite{CSE2}.

Finally, an interpretation of Cauchy coherent states in terms of the zeroth- and first-order correction to the propagation of  initial wave-packets in tilted waveguide arrays is provided.


\section*{Acknowledgments}

We thank the support of the Spanish MICINN through the project PID2022-138144NB-I00. JG thanks University of Ja\'en for an Action 1b from POAIUJA 2021-2022.

\bibliographystyle{ieeetr}



\bibliography{BibliographyCS.bib}

\end{document}